\documentclass[pra,twocolumn,showpacs,floatfix,amsfonts]{revtex4}
\usepackage{bm}
\usepackage{graphicx}
\usepackage{here}                    
\usepackage{latexsym}                
\usepackage{amsmath}
\usepackage{amssymb}  
\usepackage{dcolumn}

\newcommand{\bra}[1]{\langle #1 | \,}
\newcommand{\ket}[1]{\, | #1 \rangle}

\newcommand{\be}{\begin{equation}}
\newcommand{\ee}{\end{equation}}
\newcommand{\bea}{\begin{eqnarray}}
\newcommand{\eea}{\end{eqnarray}}

\def\unity{\openone}

\newcommand{\field}[1]{\mathbb{F}_{#1}}

\begin{document}
\title{Security bound of two-bases quantum key 
distribution protocols using qudits}
\author{Georgios M. Nikolopoulos}
\affiliation{Institut f\"ur Angewandte Physik, Technische 
Universit\"at Darmstadt, 64289 Darmstadt, Germany}
\author{Gernot Alber} 
\affiliation{Institut f\"ur Angewandte Physik, Technische 
Universit\"at Darmstadt, 64289 Darmstadt, Germany}

\date{\today}

\begin{abstract}
We investigate the security bounds of quantum cryptographic 
protocols using $d$-level systems. In particular, we focus on 
schemes that use two mutually unbiased bases, 
thus extending the BB84 quantum key distribution scheme to higher 
dimensions. Under the assumption of general coherent attacks, we 
derive an analytic expression for the ultimate upper security bound of 
such quantum cryptography schemes.  This bound is well below the 
predictions of optimal cloning machines. The possibility of 
extraction of a secret key beyond entanglement distillation is 
discussed.  In the case of qutrits we argue that any eavesdropping 
strategy is equivalent to a symmetric one. For higher dimensions such 
an equivalence is generally no longer valid. 

\end{abstract}


\maketitle

\section{Introduction}
\label{intro}
During the last two decades, several quantum key distribution 
(QKD) protocols have been proposed, which use two-level quantum 
systems (qubits) as information carriers 
\cite{BB84,qubitPRL,qubitPRA}. 
The security of these protocols against all kinds of attacks 
has been analyzed extensively and various unconditional security 
proofs have been presented \cite{LC,M,SPTKI,Lo,GLLPIM,GPTLC,GL-C}. 
From the experimental point of view, a number of 
prototypes based on qubits have been developed \cite{RMP}, while 
QKD has been successfully performed outside the laboratory at  
distances up to about $67$km using telecom 
fibers \cite{Setal,EPR}, and up to $23.4$km \cite{Open} 
through open air. 

In contrast to qubits, the use of high-dimensional quantum systems 
in quantum cryptography has attracted considerable attention only 
recently. Currently, qudits ($d$-di\-me\-nsio\-nal quantum systems) 
can be realized experimentally in several ways (including 
multiport beam splitters, bi-photons, higher-order parametric 
down-conversion and energy-time entanglement) 
\cite{QuditExp,TAZG,RMSTZG}. 
As far as QKD protocols are concerned, qudits can carry more 
information than qubits increasing thus the flux 
of information between the two legitimate users (Alice and Bob). For a 
prime power $d$ it has been demonstrated that there exist $(d+1)$ 
mutually unbiased bases. Hence, the natural extensions of the standard 
BB84 and six-state qubit-based QKD protocols to higher dimensions 
involve $2d$ and $d(d+1)$ states, respectively \cite{WF,BBRV}. 
These latter qudit-based QKD schemes are able to tolerate higher 
error rates than their qubit-based counterparts 
\cite{chau,PABM,PT,DKCK,BKBGC,AGS}.

The maximal error rate that can be tolerated by a particular QKD 
protocol (also referred to as {\em threshold disturbance}) 
quantifies the robustness 
of the protocol against a specific eavesdropping strategy, 
and depends on the algorithm that Alice and Bob are using 
for post-processing their raw key. 
In practice, nowadays secret keys can be distilled efficiently 
by means of one- or even two-way classical post-processing 
\cite{CK-BBCM,Cascade}, while advantage distillation 
protocols using two-way classical communication seem to be still 
rather inefficient \cite{Mau}. 
In principle, however, quantum distillation protocols involving 
two-way communication between Alice and Bob
[also referred to as two-way entanglement purification protocols
(EPPs)] can tolerate substantially higher error rates than 
their classical counterparts and can be applied whenever the quantum 
state shared between the two honest parties is freely entangled, 
i.e. distillable \cite{DEJ,BDSW,ADGJ,HH}. 

For $2\otimes 2$ quantum systems, non-distillability is equivalent 
to separability \cite{Letal,P-H} and thus there seems to exist 
a complete equivalence between entanglement distillation and secrecy. 
In particular, for qubit-based QKD protocols and under the assumption 
of individual attacks, it was proven recently that the extraction of a 
secret key from a quantum state is possible if and only if 
entanglement distillation is possible \cite{AMG}.  
For higher dimensions, however, the complete equivalence between 
entanglement distillation and secrecy, has been put into question 
by Horodecki {\em et al.} \cite{HHHO}, who showed that a secret key 
can, in principle, be extracted even from bound entangled states 
\cite{HHH}. 
Nevertheless, for arbitrary dimensions, {\em provable quantum 
entanglement is always a necessary precondition for secure QKD} 
\cite{CLL-AG}. 
Therefore, the natural question arises whether qudit-based 
QKD protocols can indeed go beyond entanglement distillation.  
In other words, what is the maximal error rate that can, 
in principle, be tolerated by a qudit-based QKD under the 
assumption of general coherent attacks ? 

In this paper, we address this question by focusing on qudit-based 
QKD protocols that use two mutually unbiased bases.  
Up to date, all investigations related to the security of 
such protocols have concentrated mainly on individual attacks 
(e.g. quantum cloning machines) and/or one-way post-processing 
of the raw key \cite{PABM,PT,DKCK,BKBGC,AGS}. 
Here, under the assumption of general coherent (joint) attacks, 
we show that for estimated disturbances below $(d-1)/2d$ Alice and 
Bob can be confident that they share distillable entanglement with high 
probability. 
On the other hand, an estimated disturbance above $(d-1)/2d$ does not 
enable Alice and Bob to infer that their quantum state is entangled 
({\em no provable quantum entanglement}). 
Hence, in view of the necessary precondition for secure key 
distribution \cite{CLL-AG}, our result demonstrates that 
$(d-1)/2d$ is also the ultimate threshold disturbance for the  
prepare-and-measure schemes of the protocols.
Furthermore, our result implies that, for the post-processing we 
consider throughout this work, the 
extraction of a secret key beyond entanglement distillation  
is impossible in the framework of qudit-based QKD 
protocols using two bases.

This paper is organized as follows : 
For the sake of completeness, in Sec. \ref{basics} we summarize basic 
facts which are necessary for the subsequent discussion.
In Sec. \ref{2bases} we briefly describe the prepare-and-measure 
and the enta\-ngle\-ment-based versions of the $2$-bases  
QKD protocols using qudits. 
Subsequently, Sec. \ref{D-sec} focuses on the key 
quantity of this work namely, the estimated error rate (disturbance) 
and its symmetries. 
Finally, the threshold disturbance for $2$-bases qudit-based QKD 
protocols, is derived in Sec. \ref{distil} and various examples 
are presented.  

\section{Qudits and the generalized Pauli group}
\label{basics}
Throughout this work we consider QKD protocols with qudit systems 
as information carriers. Each qudit corresponds to a $d-$dimensional 
Hilbert space $\mathbb{C}^d$ where $d=p^r$ is a 
prime power, i.e. $p$ is a prime and $r$ is an integer 
\cite{note_prime}. 
From now on all the arithmetics are performed in 
the finite (Galois) field $\field{d}$ \cite{ECC-book}.

Theoretical investigations of $d$-level quantum systems 
are performed conveniently with the help of the generalized Pauli 
group. 
For this purpose let us define the unitary operators   
\bea
{\cal X}  &=& \sum_{\alpha\in\field{d}}\ket{\alpha + 1}\bra{\alpha},\\
{\cal Z}  &=& \sum_{\alpha\in\field{d}}\omega^{{\rm tr}(\alpha)}
\ket{\alpha}\bra{\alpha},
\label{xz}
\eea
where $\omega=\exp(i2\pi/p)$ is a primitive $p$-th root of 
unity 
and 
\bea
{\rm tr}(\alpha) = \sum_{j=0}^{r-1}\alpha^{p^j}
\label{trc}
\eea
is the absolute trace of $\alpha\in\field{d}$. The states 
$\{\ket{\alpha};\alpha\in\field{d}\}$ constitute an orthonormal 
computational basis on the Hilbert space of a qudit $\mathbb{C}^d$. 
The unitary operators ${\cal X}$ and ${\cal Z}$ generate the generalized 
Pauli group  with unitary elements 
\bea
{\cal E}_{mn}= \{{\cal X}^m {\cal Z}^n: m,n\in\field{d}\}.
\label{err}
\eea
These $d^2$ unitary operators form an error group on  $\mathbb{C}^d$ \cite{ErrorGroup}, 
and are the generalizations of the Pauli operators for qubits.
In fact the indices $m$ and $n$ refer to shift and phase errors 
in the computational basis, respectively. 
Thus the generalized Pauli operators can be represented in the form 
\bea
{\cal E}_{mn} = \sum_{k\in\field{d}} \omega^{{\rm tr}(k\cdot n)}
\ket{k+m}\bra{k},
\label{err2}
\eea
with 
\bea
{\cal Z}^n{\cal X}^m&=&\omega^{{\rm tr} (m\cdot n)}{\cal X}^m
{\cal Z}^n.
\label{prop2}
\eea

Consider now a bipartite system of two qudits $A$ and $B$. It is 
not hard to show that the operators ${\cal X}_A\otimes{\cal X}_B^*$ 
and ${\cal Z}_A\otimes{\cal Z}_B^*$ 
constitute a {\em complete set of commuting operators} 
in the Hilbert space of two distinguishable qudits 
$\mathbb{C}_A^d\otimes\mathbb{C}_B^d$,  
while their simultaneous eigenstates are the $d^2$ maximally 
entangled states
\bea
\ket{\Psi_{mn}}=\frac{1}{\sqrt{d}}
\sum_{k\in\field{d}}\unity_A\ket{k_A}\otimes{\cal E}_{mn;B}\ket{k_B}, 
\label{bell-like}
\eea 
with $m,n\in\field{d}$. These states are the generalization of the 
Bell states to higher dimensions and they form an orthonormal basis in 
$\mathbb{C}_A^d\otimes\mathbb{C}_B^d$.
The singlet state $\ket{\Psi_{00}}$ is of particular interest because 
it remains invariant under any unitary transformation of the form 
${\cal U}_A\otimes{\cal U}_B^*$. 
In fact $\ket{\Psi_{00}}$ is one of the key elements of the 
entanglement-based version of the qudit cryptographic protocols 
described in the following section.

\section{Two-bases QKD protocols} 
\label{2bases}
\subsection{Mutually unbiased bases}
\label{2bases-1}
Of central importance in the context of QKD 
is the notion of mutually unbiased (maximally conjugated) bases. 
It has been demonstrated that for a prime power $d$, there exist 
$d+1$ such bases, i.e. the eigenbases of the operators
${\cal Z},\,{\cal X},{\cal XZ},\,{\cal XZ}^2,\ldots,{\cal XZ}^{d-1}$ 
\cite{WF,BBRV}.
In a qudit-based $2$-bases QKD protocol (to be referred to 
hereafter as $2d$-state protocol), Alice and Bob 
use for their purposes only two mutually unbiased bases ${\cal B}_1$ 
and ${\cal B}_2$ with $d$ basis-states each. 
Following \cite{BKBGC,AGS}, from now on the eigenbasis 
$\{\ket{k} : k\in\field{d}\}$ of the operator ${\cal Z}$ 
is chosen as the standard (computational) basis ${\cal B}_1$, 
while the second basis 
${\cal B}_2\equiv\{\ket{\bar{l}} : l\in\field{d}\}$ is chosen 
as the Fourier-dual of the computational basis, i.e.  
$\ket{\bar{l}}\equiv\sum_k{\cal H}_{lk}\ket{k}$, 
with  
\bea
{\cal H} = 
\frac{1}{\sqrt{d}}\sum_{i,j\in\field{d}}\omega^{{\rm tr}(i\cdot j)}
\ket{i}\bra{j}
\label{fourier}
\eea
denoting the discrete Fourier transformation. 
One can verify easily that ${\cal H}$ is symmetric and thus unitary, 
i.e. ${\cal H}^\dag={\cal H}^{-1}={\cal H}^*$. This property will be 
used extensively in the following  sections. 
Besides, errors in the two maximally conjugated bases are related 
via the discrete Fourier transform, i.e.   
\bea
{\cal H}^\dag {\cal E}_{mn} {\cal H} = 
\omega^{-{\rm tr}(m\cdot n)}{\cal E}_{nm}^*.
\label{errorsMC}
\eea
In other words, shift errors in the computational basis become phase 
errors in the complementary basis and vice-versa. 

\subsection{Prepare-and-measure QKD scheme}
\label{2bases-2}
In a typical $2d$-state prepare-and-measure scheme 
Alice sends to Bob a sequence of qudits each of which is randomly 
prepared in one of the $2d$ non-orthogonal basis-states 
$\{\ket{k}\}$ or $\{\ket{\bar{l}}\}$.  
Bob measures each received particle randomly in 
${\cal B}_1$ or ${\cal B}_2$. 
After the distribution stage, Alice and Bob agree on a random 
permutation of their data and publicly discuss the bases chosen, 
discarding all the dits where they have selected different bases 
(sifting procedure). 
Subsequently, they randomly select a sufficient number of dits \cite{half}
from the remaining random sifted key and determine their error probability. 
If, as a result of a noisy quantum channel or of an eavesdropper, the  
estimated disturbance is too high the protocol is aborted.
Otherwise, Alice and Bob perform  error correction and privacy 
amplification with one- or two-way classical communication, in order 
to obtain a smaller number of secret and perfectly correlated random 
dits \cite{GL-C,chau,BKBGC,AGS,CK-BBCM,Cascade,Mau}. 

\subsection{Entanglement-based  QKD scheme}
\label{2bases-3}
From the point of view of an arbitrarily powerful eavesdropper the 
above prepare-and-measure scheme is equivalent to an 
entanglement-based QKD protocol \cite{chau,AGS,BBM}.
In this latter form of the protocol Alice prepares each of $2N$ 
entangled-qudit pairs in the maximally entangled 
state 
\bea
\ket{\Psi_{00}}=\frac{1}{\sqrt{d}}\sum_{k\in\field{d}}
\ket{k_A}\otimes\ket{k_B},
\eea
where the subscripts $A,B$ refer to Alice and Bob, respectively.  
Alice uses for her purposes the set of bases 
$\{{\cal B}_1, {\cal B}_2\}$ whereas Bob uses the set 
$\{{\cal B}_1, {\cal B}_2^*\}$, 
where ${\cal B}_2^*\equiv\{{\cal H}^*\ket{k} : k\in\field{d}\}$ 
\cite{chau,AGS}. 

More precisely, Alice keeps half of each pair and submits the other 
half to Bob after having applied a random unitary transformation 
chosen from the set $\{\unity, {\cal H}\}$. 
As soon as Bob acknowledges the reception of all the particles, 
Alice reveals the sequence of operations she performed on 
the transmitted qudits and Bob undoes all of them, i.e. he applies 
$\unity$ or ${\cal H}^{-1}$ on each qudit separately. 
Thus, at this point, in an ideal system 
Alice and Bob would share $2N$ qudit-pairs in the state 
$\ket{\Psi_{00}}^{\otimes 2N}$.   
However, in real systems, due to noise and/or eavesdropping all 
the $2N$ entangled-qudit pairs will be corrupted. 
In order to ensure secret key distribution  Alice and Bob 
{\em permute randomly} all the pairs before doing any other 
operations \cite{GL-C}.
In this way, any influence of the eavesdropper (from now on we assume 
that all the noise in the channel is due to eavesdropping) 
is equally distributed among all the pairs. 

The next step of the protocol now involves a verification test 
which will determine whether the protocol should be aborted or not. 
More precisely, Alice and Bob randomly select a number of pairs 
(say $N_{\rm c}$) 
\cite{half} as check pairs and measure each one of them 
{\em separately} along the standard (computational) basis.  
They compare 
their results publicly thus estimating the average error rate during 
the transmission. 
After the verification test all the check pairs are dismissed and,
if the estimated error rate is too high the protocol is aborted.
Otherwise, Alice and Bob apply an appropriate EPP with classical 
one- or two-way communication \cite{chau,DEJ,BDSW,ADGJ,HH} on the 
remaining $2N-N_{\rm c}$ pairs, in order to distill a smaller number of almost 
pure entangled-qudit pairs. Finally, measuring these  almost perfectly
entangled qudit pairs in a common basis, Alice and Bob obtain a 
secret random key, about which an adversary has negligible 
information. In our subsequent treatment we focus on the 
entanglement-based version of the $2d$-state QKD protocol. 

\section{Estimated disturbance and symmetries} 
\label{D-sec}
The verification test performed by Alice and Bob 
immediately after the transmission stage is perhaps the most 
crucial stage of the $2$-bases QKD protocol and its success relies on 
the ``commuting-observables'' idea \cite{LC}. More precisely, the fact 
that all the operations performed in a typical EPP 
commute with a Bell measurement allows one to reduce any quantum 
eavesdropping attack to a classical probabilistic cheating strategy 
\cite{Lo,LC,chau,GL-C}.  

During the verification test Alice and Bob focus 
on the parity of their outcomes.
Moreover, note that for the check pairs where Alice and Bob have 
performed ${\cal H}$ and ${\cal H}^{-1}$ respectively, the 
measurements are effectively performed in the complementary 
${\cal B}_2$ basis rather than the standard basis ${\cal B}_1$ 
\cite{SPTKI}. 
Thus, given the unitarity of ${\cal H}$ and the invariance of 
$\ket{\Psi_{00}}$ under any unitary transformation of the form 
${\cal U}_A\otimes{\cal U}_B^*$, the average estimated disturbance 
(error rate) is given by
\begin{widetext}
\begin{eqnarray}
D = \frac{1}{2 N_{\rm c}}\sum_{b=0,1}\sum_{j_i=1}^{N_{\rm c}}
{\rm Tr}_{A,B}\left \{
\left [\left ({\cal H}_A^{b\dag}\otimes
{\cal H}_B^b\right )
{\cal P}\left ( 
{\cal H}_A^b
\otimes{\cal H}_B^{b\dag}\right )
\right ]_{j_i}\rho_{AB}
\right \},
\label{QBER2-1}
\end{eqnarray}
\end{widetext}
where $\rho_{AB}$ denotes the reduced density operator of 
Alice and Bob for all $2N$ pairs. 
The index $j_i$ indicates that the corresponding physical observable 
refers to the $j_i$-th randomly selected qudit-pair. 
In particular, the projection operator entering Eq. (\ref{QBER2-1}) 
is given by 
\bea
{\cal P}_{j_i} \equiv \sum_{l\in\field{d}}
\sum_{k\in\field{d}^*}\ket{l_A,(l+k)_B}
\bra{l_A,(l+k)_B},
\label{Proj1}
\eea
where $\field{d}^*$ denotes the set of all nonzero elements in the 
field $\field{d}$  \cite{notation}. In other words, the inner summation in 
(\ref{Proj1}) is performed over 
all the nonzero elements of the finite field $\field{d}$, 
such that $(l+k)_B\neq l_A$.
Moreover, the powers of the discrete Fourier transformation 
${\cal H}^b$, with $b\in\{0,1\}$, in Eq. (\ref{QBER2-1}) reflect the fact that
the errors in the sifted key originate from measurements in 
both complementary bases which have been selected randomly by Alice 
and Bob with equal probabilities.
One can easily verify that all the measurements performed  
during the verification test are equivalent to Bell measurements. 
Indeed, using the definition of the Bell states (\ref{bell-like}) the 
projector ${\cal P}_{j_i}$ can be written in the form  
\bea
{\cal P}_{j_i} = \sum_{m,n\in\field{d}}(1-\delta_{m,0})\ket{\Psi_{mn}}
\bra{\Psi_{mn}},
\label{Proj2}
\eea
where $\delta_{m,0}$ is the Kronecker delta \cite{notation}. 
This last relation indicates that the verification test performed by 
Alice and Bob is nothing else than a quality-check test of the  
fidelity of the $2N$ pairs with respect to the ideal state 
$\ket{\Psi_{00}}^{\otimes 2N}$. Hence, classical sampling theory  
can be applied for the estimation of the average error rate and 
the establishment of confidence levels \cite{Lo,LC,chau,GL-C}. 

We can simplify further our discussion by taking into 
account the symmetry of the QKD protocol under any permutation 
of the pairs. As we discussed earlier, a random permutation of all 
the pairs at the beginning of the entanglement-based protocols 
ensures a homogeneous distribution of the errors introduced by a 
potential eavesdropper (Eve) over all the qudit pairs \cite{GL-C}. 
This is equivalent to saying that the eavesdropping attack is 
symmetric on all the pairs, and such a symmetrization argument 
is one of the key elements of various unconditional security proofs 
\cite{SPTKI,Lo,GL-C,chau}. 
Indeed, Eve does not know in advance which of the qudit-pairs will be 
used for quality checks and which qudit-pairs will contribute to 
the final key. Hence, she is not able to treat them 
differently and the check pairs constitute a classical random 
sample of all the pairs. 

Invariance of the eavesdropping attack under 
any permutation of the pairs implies that all the reduced density 
operators describing the state of each pair shared between Alice 
and Bob are equal, i.e.
\begin{eqnarray}
\rho_{AB}^{(1)} &=& \rho_{AB}^{(2)} =\cdots = \rho_{AB}^{(2N)},
\label{homogen}
\end{eqnarray}
where the reduced density operator of Alice's and Bob' s $k$-th 
pair is denoted by
$\rho_{AB}^{(k)} = {\rm Tr}_{AB}^{(\not k)}(\rho_{AB})$, with 
${\rm Tr}_{AB}^{(\not k)}$ indicating the tracing (averaging) procedure
over all the qudit-pairs except the $k$-th one. It should be stressed 
that Eq. (\ref{homogen}) does not at all imply that
the overall reduced density operator $\rho_{AB}$ of the $2N$ 
pairs itself, is  a product 
state of all the reduced pair states $\rho_{AB}^{(k)}$. 
On the contrary, $\rho_{AB}$ is expected to have a complicated 
structure as it includes all the effects arising from a general
coherent (joint) attack of a possible eavesdropper.

In view of Eq. (\ref{homogen}), the average disturbance defined in 
Eqs. (\ref{QBER2-1}) is 
determined by the average error probability of an arbitrary qudit 
pair, say the pair $j_1$, i.e.
\begin{widetext} 
\begin{eqnarray}
D = \frac{1}{2}\sum_{b=0,1}
{\rm Tr}_{A,B}^{(j_1)}\left \{
\left [ 
\left ({\cal H}_A^{b\dag} 
\otimes{\cal H}_B^b\right ) 
{\cal P}\left ({\cal H}_A^b\otimes
{\cal H}_B^{b\dag}\right )
\right]_{j_1}\rho_{AB}^{(j_1)}
\right \},
\label{QBER2-2}
\end{eqnarray}
\end{widetext}
where ${\rm Tr}_{A,B}^{(j_1)}$ denotes the 
tracing procedure over the $j_1$-th qudit pair of Alice and Bob. 
In other words, the reduced single-pair state $\rho_{AB}^{(j_1)}$
contains all the information about the noisy quantum channel and a 
possible general coherent attack
by an eavesdropper, which is relevant for the evaluation of 
the error rate.
In particular, this implies that an arbitrary joint eavesdropping 
attack which gives rise to 
a particular state $\rho_{AB}$ obeying Eq. (\ref{homogen}) is 
indistinguishable, from the 
point of view of the estimated disturbance, from
a corresponding collective attack which addresses each qudit 
individually and results in the $2N$-pair 
state of the form $\bigotimes_{j=1}^{2N}\rho_{AB}^{(j)}$, 
for example.

According to Eqs. (\ref{Proj1}) and (\ref{QBER2-2}) the average 
estimated disturbance is invariant under the transformations
\begin{subequations}
\label{SymmG1}
\bea
(l,b) &\to& (l + m,b),\\ 
(l,b) &\to& (l,b \oplus 1),
\eea
\end{subequations}
with $ m\in\field{d}$, while $\oplus$ denotes addition modulo $2$.
This invariance implies that there are various reduced density 
operators of the $j_1$-th qudit pair, which all give rise
to the same observed value of the average disturbance. 
This can be seen from Eq. (\ref{errorsMC})
which implies elementary relations of the form 
\begin{widetext}
\bea
{\cal E}_{mn}{\cal H}^b
\ket{j}\bra{j}({\cal H}^b)^\dag
{\cal E}_{mn}^\dag =
{\cal H}^b \ket{j+bn+(1-b)m}\bra{j+bn+(1-b)m} {\cal H}^{b\dag}.
\label{simpleG}
\eea
\end{widetext}
Together with the invariance of $D$ under the 
transformations (\ref{SymmG1}), these elementary relations imply that
the reduced operators $\rho_{AB}^{(j_1)}$ and the symmetrized state 
\begin{eqnarray}
\tilde{\rho}_{AB}^{(j_1)} &=&\frac{1}{4d^2}
\sum_{g\in{\cal G}_1, h\in {\cal G}_2} 
U(h)U(g)\rho_{AB}^{(j_1)}U(g)^{\dagger}U(h)^{\dagger} 
\label{rhotildeG2}
\end{eqnarray}
give rise to the same value of $D$. 
Thereby, the unitary operators
\begin{eqnarray}
U(g_{mn}) &=& {\cal E}_{mn;A}\otimes {\cal E}_{mn;B}^*,
\label{EU}
\end{eqnarray}
\begin{eqnarray}
U(h_1) &=& \unity_A\otimes \unity_B,\quad\,\,
U(h_3) = ({\cal H}_A\otimes {\cal H}_B^*)^2,\nonumber\\ 
U(h_2) &=& {\cal H}_A\otimes {\cal H}_B^*,\quad
U(h_4) = ({\cal H}_A\otimes {\cal H}_B^*)^3,
\label{HU}
\end{eqnarray}
have been introduced, which
form  unitary representations of two discrete Abelian groups
${\cal G}_1 =\{g_{00},g_{01},\ldots\}$ and 
${\cal G}_2 =\{h_1,h_2,h_3,h_4\}$.
The key point is now that, invariance of $\tilde{\rho}_{AB}^{(j_1)}$ 
under  both of these groups is induced by the symmetry 
transformations (\ref{SymmG1}) which leave $D$ invariant.

\section{Entanglement distillation and secret key}
\label{distil}
Having exploited the symmetries underlying the estimated disturbance,  
in this section we estimate the threshold 
disturbance that can, in principle, be tolerated by any 
$2d$-state QKD protocol, under the assumption of arbitrary 
coherent (joint) attacks. 
To this end, we make use of the {\em necessary precondition} for 
secret 
key distillation that is, the correlations established between Alice 
and Bob
during the state distribution cannot be explained by a separable
state \cite{CLL-AG}. 

Throughout this work, we consider that Alice and
Bob focus on the sifted key during the post processing
(i.e., they discard immediately all the polarization data for which
they have used different bases) and that they treat each pair
independently. 
Thus, according to the aforementioned precondition, given a particular 
value of the estimated disturbance $D$, the task of Alice and Bob 
is to infer whether their correlations may have originated from a 
separable state or not. 
So, {\em our aim is to estimate the threshold disturbance 
$D_{\rm th}$ such that for any $D<D_{th}$ Alice and Bob share 
provable entanglement with certainty}. To this end, we proceed as 
follows :  Firstly, we estimate the regime of disturbances for which 
Alice and Bob share distillable entanglement. Secondly, we demonstrate 
that for the remaining regime of disturbances the correlations shared 
between Alice and Bob can always be described by a separable state. 

\subsection{Threshold disturbance} 
\label{distil-1}
Adopting the entanglement-based version of the protocol defined in 
Sec. \ref{2bases-3}, let us estimate the regime of disturbances for 
which Alice and Bob share free entanglement. 
From the symmetries underlying the observed average error rate   
and in particular from Eq. (\ref{rhotildeG2}) we have that 
the density operator $\rho_{AB}^{(j_1)}$ is freely entangled 
if $\tilde{\rho}_{AB}^{(j_1)}$ is freely entangled, as both states are 
related by local unitary operations and convex summation. 
Hence, to determine the values of the disturbance for which the real 
state $\rho_{AB}^{(j_1)}$ is distillable, it suffices to determine the 
disturbances for which the most general two-qubit state 
$\tilde{\rho}_{AB}^{(j_1)}$  
(which is invariant under the discrete Abelian groups ${\cal G}_1 $ 
and ${\cal G}_2$) is distillable. 

We already know that the operators $U(g_{10})\equiv{\cal X}_A
\otimes{\cal X}_B^*$ 
and $U(g_{01})\equiv{\cal Z}_A\otimes{\cal Z}_B^*$ of the group 
${\cal G}_1$ constitute a {\em complete set of commuting operators} 
in  $\mathbb{C}_A^d\otimes\mathbb{C}_B^d$, while their simultaneous 
eigenstates are the $d^2$ maximally entangled states defined in 
Eq. (\ref{bell-like}).
Thus, the most general two-qudit state which is invariant under the 
Abelian group ${\cal G}_1$ is given, by a convex sum of 
all $\ket{\Psi_{mn}}$, 
i.e. 
\bea
{\tilde \rho}_{AB}^{(j_1)}= \sum_{m,n\in\field{d}}
\lambda_{mn}\ket{\Psi_{mn}}\bra{\Psi_{mn}},
\label{rhoAB-Bell-G}
\eea
where the non-negative parameters $\lambda_{m n}$ 
have to fulfill the normalization condition 
\bea 
\sum_{m,n\in\field{d}}\lambda_{mn} = 1.
\label{normG}
\eea 
Moreover, the operations  
${\cal H}_A\otimes {\cal H}_B^*$,  
$({\cal H}_A\otimes {\cal H}_B^*)^2$ and 
$({\cal H}_A\otimes {\cal H}_B^*)^3$  
transform Bell states into other Bell states. 
Thus, additional invariance of the quantum state 
(\ref{rhoAB-Bell-G}) under the discrete group ${\cal G}_2$ implies 
that 
\bea
\lambda_{m,n}=\lambda_{n,d-m}=\lambda_{d-m,d-n}
=\lambda_{d-n,m}. 
\label{lambdas-2}
\eea
As a consequence of Eq. (\ref{lambdas-2}) there are 
different sets of identical parameters $\lambda_{mn}$. 
Each set $j$ contains four members $\eta_j$ unless the chain 
(\ref{lambdas-2}) 
is truncated. The latter case occurs for $d-m=m$ and $d-n=n$,  
i.e. for $m,n\in\{0,d/2\}$. More precisely, the sets $j$ with 
$m=n\in\{0,d/2\}$ contain one eigenvalue $\xi_j$ each, whereas 
the set with  
$m\neq n\in\{0,d/2\}$ has two equal eigenvalues denoted by $\zeta$. 
From now on we distinguish between even and odd dimensions 
$d$. All the sets for both cases as well as their notation are 
summarized in Table~\ref{tab:table1}.

\begin{table}
\caption{
The notation of the sets and the number of eigenvalues per set 
for even and odd dimensions.}
\label{tab:table1}
\begin{ruledtabular}
\begin{tabular}{cccc}
members per set &\multicolumn{2}{c}{number of sets} & notation\\
 & Even $d$ & Odd $d$ & \\
\hline
1 & 2 &  1 & $\xi_j$\\
2 & 1 & 0 &$\zeta$ \\
4 & $(d^2-4)/4$ & $(d^2-1)/4$  & $\eta_j$ \\
\end{tabular}
\end{ruledtabular}
\end{table}

Given the various sets of eigenvalues, the normalization 
condition (\ref{normG}) now reads
\begin{subequations}
\bea
{\rm Odd }\,\, d&:&\quad \xi_0+4\sum_{j=1}^{\eta_{odd}}\eta_j=1,\nonumber\\
{\rm Even }\,\, d&:&\quad \xi_0+\xi_1+2\zeta+4\sum_{j=1}^{\eta_{even}}\eta_j=1,
\nonumber
\eea
\end{subequations}
where in both cases the index $j$ runs over all the possible 
4-member groups i.e., $\eta_{odd}\equiv(d^2-1)/4$ and 
$\eta_{even}\equiv(d^2-4)/4$ (see Table~\ref{tab:table1}). 
Similarly, using Eqs. (\ref{Proj2}), (\ref{QBER2-2}) and 
(\ref{rhoAB-Bell-G}), 
the estimated average disturbance can be expressed in the form  
\begin{subequations} 
\bea
{\rm Odd}\,\, d &:& \quad
D=2\sum_{j=1}^{\lfloor d/2\rfloor}\eta_j+
4\sum_{j=\lfloor d/2\rfloor+1}^{\eta_{odd}}\eta_j,\nonumber\\
{\rm Even}\,\, d &:& \quad
D=\xi_1+\zeta+2\sum_{j=1}^{d/2-1}\eta_j+
4\sum_{j=d/2}^{\eta_{even}}\eta_j,\nonumber
\label{D-general}
\eea
\end{subequations}
with $\lfloor x\rfloor$ denoting the largest integer not greater than $x$, 
while all the parameters (disturbance and eigenvalues) are real-valued and 
non-negative, i.e. 
\bea
0\leq D,\xi_j,\zeta,\eta_j\leq 1.\nonumber
\eea

Let us evaluate now the disturbances for which the state 
${\tilde \rho_{AB}}^{(j_1)}$ is distillable.  
According to the reduction criterion \cite{HH}, if 
${\tilde \rho_{AB}}^{(j_1)}$ is separable, then   
\bea
{\tilde \rho}_A^{(j_1)}\otimes\unity_B-
{\tilde \rho_{AB}}^{(j_1)}\geq 0
\label{red}
\eea 
(and also $\unity_A\otimes{\tilde \rho}_B^{(j_1)}-
{\tilde \rho}_{AB}^{(j_1)}\geq 0$), with 
${\tilde \rho}_A^{(j_1)}\equiv {\rm Tr}_B 
[{\tilde \rho}_{AB}^{(j_1)}]$. 
Using the explicit form of ${\tilde \rho_{AB}}^{(j_1)}$ given by 
Eq. (\ref{rhoAB-Bell-G}) we have 
${\tilde \rho}_A^{(j_1)}={\tilde \rho}_B^{(j_1)}=\unity_d/d$, 
where $\unity_d$ denotes the unit operator in 
$\mathbb{C}_{A(B)}^d$. Thus inequality (\ref{red}) 
reads
\bea
\sum_{m,n\in\field{d}}
\left (\frac{1}{d}-\lambda_{mn}\right )\ket{\Psi_{mn}}\bra{\Psi_{mn}}
\geq0.
\label{dstlCnst}
\eea
Violation of the above inequality (\ref{dstlCnst}) for any of the 
eigenvalues $\lambda_{mn}$, i.e. 
\bea
\lambda_{mn}> \frac{1}{d}, 
\label{distilConst1}
\eea
is {\em sufficient}  for distillability of the 
entanglement of ${\tilde \rho}_{AB}^{(j_1)}$ and implies violation 
of the Peres criterion (i.e., a non-positive partial transpose) 
for this state \cite{HH,Letal}. 
In particular, as long as the fidelity 
$f$ of ${\tilde \rho}_{AB}^{(j_1)}$ with respect to $\ket{\Psi_{00}}$ 
satisfies  
\bea
f\equiv\bra{\Psi_{00}}{\tilde \rho}_{AB}^{(j_1)}\ket{\Psi_{00}}>
\frac{1}{d},
\label{fidel}
\eea 
the state  can be distilled with the help of unitary twirling 
operations ${\cal U}_A\otimes{\cal U}_B^*$ which leave $f$ invariant 
\cite{HH}.
In our case, using Eqs. (\ref{rhoAB-Bell-G}) and (\ref{lambdas-2})  
the distillability condition (\ref{fidel}) reads 
$\xi_0 > 1/d$ or equivalently  
\begin{subequations}
\bea
{\rm Odd}\quad d&:&\quad D < D_0
+2\sum_{j=\lfloor d/2\rfloor+1}^{\eta_{odd}}\eta_j,\nonumber
\\
{\rm Even}\quad d&:&\quad D< D_0+\frac{1}{2}\xi_1+
2\sum_{j=d/2}^{\eta_{even}}\eta_j,\nonumber
\eea
\end{subequations}
where 
\bea
D_0\equiv \frac{d-1}{2d}.
\label{d0-def}
\eea

According to these last inequalities, and given the fact that 
$\xi_j,\zeta,\eta_j\geq 0$, the threshold disturbance $D_{\rm th}$ 
for entanglement distillation at any dimension satisfies the 
inequality   
\bea
D_{\rm th}\geq D_0,
\label{distilConst3}
\eea 
with $D_0$ given by (\ref{d0-def}).
For any $D<D_{\rm th}$, the symmetrized state 
${\tilde \rho}_{AB}^{(j_1)}$ is always distillable 
(i.e., freely entangled). Given that $\rho_{AB}^{(j_1)}$ and 
${\tilde \rho}_{AB}^{(j_1)}$ 
are related via local operarations and convex summation, the original 
state $\rho_{AB}^{(j_1)}$ must also be distillable in the same regime 
of disturbances. 

Nevertheless, the fact that inequality (\ref{fidel}) is not satisfied 
for $D\geq D_{\rm th}$, does not necessarily imply that the state 
${\tilde \rho}_{AB}^{(j_1)}$ is not at all distillable for 
$D\geq D_{\rm th}$. 
For instance, there might exist another eigenvalue $\lambda_{mn}$ 
(and not $\xi_0$)  
which satisfies inequality (\ref{distilConst1}) [i.e., it violates 
inequality (\ref{red})] and this fact, according to the the reduction 
criterion, is also sufficient for distillability of 
${\tilde \rho}_{AB}^{(j_1)}$ \cite{HH}. 
Hence, we must now evaluate the precise value of the threshold 
disturbance $D_{\rm th}$. 

One way to prove that strict equality holds in 
Eq. (\ref{distilConst3}) for any $2d$-state QKD protocol is to 
demonstrate that for $D \geq D_0$, there always exist separable 
states which can describe Alice's and Bob's correlations and 
simultaneously are indistinguishable from the real bipartite state 
$\rho_{AB}^{(j_1)}$. 
To this end, let us focus on bipartite Bell-diagonal states 
i.e., states which 
can be written in the form (\ref{rhoAB-Bell-G}) and consider the 
following particularly simple family of such separable 
states 
\begin{widetext}
\bea
\sigma_{AB}(D)&=&y\unity_{d^2}+d|x-y|\sum_{k\in\field{d}} 
\frac{(\ket{k_A}\bra{k_A})\otimes
(\ket{k_B}\bra{k_B})}{d}+d|x-y|\sum_{i\in\field{d}}
\left ({\tilde \sigma}_{A}^{(i)} 
\otimes {\tilde \sigma}_{B}^{(i)}\right).
\label{sep}
\eea 
\end{widetext}
Thereby 
\bea 
x &=& \frac{1+d(d-2)(1-D)}{d^2(d-1)},\nonumber \\
y &=& \frac{1+d(-1+2D)}{d^2(d-1)}, \nonumber
\eea 
and 
\bea
{\tilde \sigma}_{C}^{(i)} &=& 
\frac{1}{d}\sum_{k\in\field{d}}\ket{k_{C}}\bra{(k+i)_{C}}, 
\nonumber
\eea 
while $\unity_{d^2}$ denotes the unit operator in 
$\mathbb{C}_A^d\otimes\mathbb{C}_B^d$.

This family is parametrized by the estimated average disturbance $D$ 
detected by Alice and Bob and is valid for 
\bea
\frac{d-1}{2d}=D_0\leq D\leq \frac{2d-1}{2d}. 
\label{interval}
\eea 
Moreover, any separable state which belongs to this family is 
indistinguishable, from the point of view of the estimated error rate, 
from the real state shared between Alice and Bob. 
In other words, whenever the detected disturbance $D$ is within the 
interval (\ref{interval}), the correlations shared between Alice and 
Bob can be 
very well described in the framework of the family of seprable states 
$\sigma_{AB}(D)$. In such a case, the necessary precondition for 
secret key distillation is not met for disturbances within this 
regime, so that the protocol must be aborted. 

So, we have proved that strict equality holds in 
(\ref{distilConst3}) and thus, from Alice's and Bob's point of view, 
the threshold disturbance for entanglement distillation in the context 
of entanglement-based $2d$-state QKD protocols is 
\bea
D_{\rm th}=\frac{d-1}{2d}.
\label{distilConst4}
\eea  
In particular, if the detected average disturbance is below this 
threshold, the two legitimate users can be assured that 
they share freely entangled qudit-pairs with high probability. 
In other words, under the assumption of general coherent attacks, 
for $D<D_{\rm th}$ Alice and Bob are always able to extract a secret 
key by application of a two-way EPP which purifies towards the 
maximally entangled state $\ket{\Psi_{00}}$.
On the other hand, an estimated disturbance above 
$D_{\rm th}=(d-1)/2d$, does not allow Alice and Bob to infer whether 
the state they share is entangled or not. 
In particular, we have seen that there is at least one simple 
family of separable states which can describe Alice's and Bob's 
correlations up to high error rates of magnitude $(2d-1)/2d$. 
Finally, note that for $d=2$ we reveal the threshold disturbance 
for the standard BB84 QKD protocol, that is 
$D_{\rm th}=1/4$ \cite{NA}. 
Moreover, $D_{\rm th}\to 1/2$ for $d\to\infty$ reflecting the 
possible advantage of using higher-dimensional quantum systems as 
information carriers in quantum cryptography. 

\begin{figure}[t]
\centerline{\includegraphics[width=8.0cm]{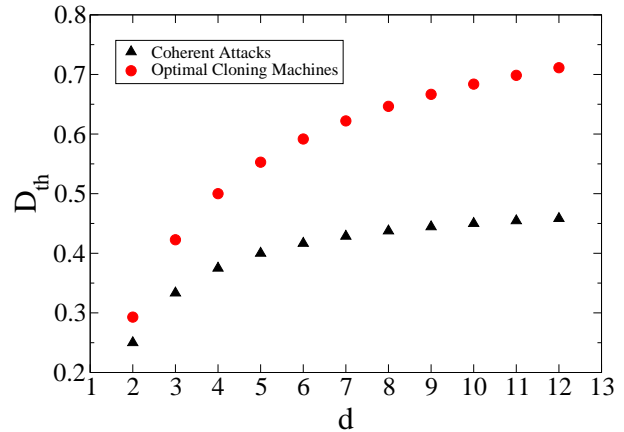}}
\caption{(Color online) $2d$-state QKD protocols : 
The threshold disturbance for entanglement distillation as 
a function of dimension. The triangles refer to 
Eq. (\ref{distilConst4}) 
and arbitrary coherent attacks whereas the circles correspond to 
Eq. (\ref{distilConstOC}) and optimal cloning machines.}
\label{Dth:fig}   
\end{figure}

In view of the {\em necessary precondition} for secret key 
distillation \cite{CLL-AG}, our results imply that $D_{\rm th}$ is 
also the ultimate upper security bound of any $2d$-state 
prepare-and-measure QKD protocol. 
Nevertheless, the details of a particular prepare-and-measure scheme 
(that is the error correction and privacy amplification protocols 
required) which will be capable of meeting this upper security 
bound remain an open question. In fact one has to specify a classical 
distillation (post-processing) protocol which has the same bounds of 
tolerable noise as quantum distillation protocols.  
It is worth mentioning, however, that the security bound 
(\ref{distilConst4}) relies on certain conditions. 
In particular, it relies on the complete omission 
of any polarization data from the raw key that involve different bases
for Alice and Bob, as well as on the individual manipulation
of each pair during the post-processing. 
If some of these conditions are changed, also the threshold
disturbance may change. 

Recently, under the same conditions, Acin {\em et al.} \cite{AGS}
derived another bound for entanglement distillation, namely 
\bea
D_{\rm th}^{\rm (CM)}=1-\frac{1}{\sqrt{d}}.
\label{distilConstOC}
\eea 
As depicted in Fig.~\ref{Dth:fig}, $D_{\rm th}^{\rm (CM)}$  
is well above the threshold we have derived in this work for any 
dimension of the information carriers. 
The reason is basically that $D_{\rm th}^{\rm (CM)}$ has been obtained 
under the additional assumption that Eve is restricted to so-called 
``optimal incoherent attacks''.  These attacks rely on cloning 
machines and maximize Eve's information gain. One can easily verify, 
for example, that the class of separable states (\ref{sep}) is not 
optimal (in the sense of \cite{AGS}). In our work we allow for 
arbitrary eavesdropping attacks and thus we have demonstrated that 
the distillation of a secret key  for disturbances above $D_{\rm th}$ 
is impossible.  
So, although the incoherent attacks considered in \cite{AGS} are 
optimal with respect to the information gain of an eavesdropper, 
they are not able to disentangle Alice and Bob at the lowest possible 
disturbance. 
The cost of information loss that Eve has to accept by employing an 
attack that disentangles Alice and Bob at each particular disturbance 
above $D_{\rm th}$ remains an open question. 
Clearly, to this end one has to consider in detail the eavesdropping 
attack and this is beyond the purpose of this work.

A further issue ought to be brought up here in connection with the 
existence of bound entanglement. For $2\otimes 2$ systems 
(i.e., for the BB84 QKD protocol) non-distillability is equivalent to 
separability \cite{P-H} and this fact seems to lead to a complete 
equivalence between entanglement distillation and secrecy \cite{AMG}. 
However, for higher dimensions the situation is more involved due to 
the existence of bound entangled states with positive or non-positive 
partial transpose \cite{Letal,HHH}. Moreover, in a recent work  
\cite{HHHO} Horedecki {\em et al.} showed that a secret 
key can be distilled even from bound entangled states. 
As a consequence, a qudit-based (with $d>2$) QKD scheme could, in 
principle, go beyond entanglement distillation.  However, this does 
not seem to be the case for the post processing 
and the protocols we consider throughout this work. 

Indeed, for $D<D_{\rm th}$ we have seen that the state shared 
between Alice and Bob is always distillable, i.e. it is freely 
entangled. Bound entangled states are expected to exist for 
$D\geq D_{\rm th}$ and this is 
precisely the regime of parameters where the ideas presented in 
\cite{HHHO} can be used for the extraction of a secret key beyond 
entanglement distillation. 
We have demonstrated, however, that an eavesdropper is always able to 
break any entanglement between Alice and Bob for $D\geq D_{\rm th}$ 
without being detected,    
by preparing, for example, a separable state from the family 
$\sigma_{AB}(D)$. 
As a consequence, according to \cite{CLL-AG}, the 
protocol must be aborted at $D=D_{\rm th}$. 
Under these circumstances, the extraction of a secret key beyond 
entanglement distillation seems to be practically impossible. 
The reason is basically that, based on the estimated error rate, 
Alice and Bob are incapable of verifying whether they share a 
separable state or not for disturbances above $D\geq D_{\rm th}$. 
Alice and Bob can improve their situation only if they do not restrict 
themselves to the sifted data only. In particular, constructing 
appropriate entanglement witnesses from their raw data \cite{CLL-AG}, 
Alice and Bob can verify whether they share a separable state or not, 
even for  $D\geq D_{\rm th}$. 

Closing this section, let us briefly compare the performance of 
two different realizations of a six-state QKD protocol, namely 
a  $3$-bases scheme using qubits and a qutrit-based scheme 
using $2$ out of $4$ mutually unbiased bases. 
In principle, both protocols can tolerate 
precisely the same error rate, that is $1/3$. 
Nevertheless, the qutrit-based protocol offers a higher yield since  
$1/2$ of the transmissions pass the sifting procedure 
(compared to $1/3$ for the qubit-based protocol).
Thus, although both six-state protocols appear to be equally secure, 
the qutrit based scheme seems to be more efficient. 

\subsection{Examples} 
\label{distil-3} 
So far, our discussion involved arbitrary dimensions and general 
coherent attacks. 
For the sake of illustration, in this subsection we briefly discuss 
low dimensions (i.e., $d=2,3$) as well as symmetric (isotropic) 
channels \cite{BKBGC,AGS} and arbitrary dimensions. 
In particular, we present evidence of the fact 
that for $d=3$ any eavesdropping strategy is equivalent to a 
symmetric one. 
However, for $d>3$ this equivalence does not seem to exist anymore.  
Moreover, we present numerical results for $d=3,4$ and $5$, 
verifying the security bounds derived in the previous subsection. 

\subsubsection{Qubits} 
As a consequence of Eq. (\ref{lambdas-2}), for $d=2$ there are 
three different eigenvalues entering Eq. (\ref{rhoAB-Bell-G}). 
So, in a matrix form we may write    
\bea
\lambda_{mn}=\left (
\begin{array}{ccc}
u\quad & x \\
x\quad & y\\ 
\end{array}
\right ), 
\label{lambdas2}
\eea 
with the eigenvalues $u,x$ and $y$ satisfying the normalization 
condition $u+2x+y=1$. 
In this notation, the estimated disturbance can be expressed in the 
form $D = x + y$. 
One can easily verify that the state $\tilde{\rho}_{AB}^{(j_1)}$ 
is entangled for $1/4<D$ and $D>3/4$ \cite{NA}. Moreover, for   
$1/4\leq D\leq 3/4$ the state $\tilde{\rho}_{AB}^{(j_1)}$ is always 
separable and indistinguishable 
(as far as the estimated disturbance is 
concerned) from the real state $\rho_{AB}^{(j_1)}$ shared 
between Alice and Bob. 

\subsubsection{Qutrits} 
In analogy to qubits, applying Eq. (\ref{lambdas-2}) for $d=3$ and 
without any additional assumptions one finds that there are 
three different eigenvalues entering 
Eq. (\ref{rhoAB-Bell-G}). In particular, the matrix of eigenvalues 
reads
\bea
\lambda_{mn}=\left (
\begin{array}{ccc}
u\quad & x\quad & x \\
x\quad & y\quad & y \\
x\quad & y\quad & y 
\end{array}
\right ), 
\label{lambdas3}
\eea 
while the average estimated disturbance is of the form   
$D = 2x + 4y$. Hence, taking into account the normalization 
condition $u+4(x+y)=1$, we have two real-valued and non-negative 
independent parameters in the problem. 
Moreover, the partial transpose of 
$\tilde{\rho}_{AB}^{(j_1)}$ is block-diagonal with 
all three blocks being identical and equal to 
\bea
M_3=\frac{1}{3}\left (
\begin{array}{ccc}
u+2x & x-y & x-y\\
x-y & x+2y & u-x \\
x-y & u-x & x+2y 
\end{array}
\right ).
\nonumber
\eea 
Hence, the following two eigenvalues 
\begin{subequations}   
\bea
\nu_1&=&\frac{1}{3}\left(-u+2x+2y\right),\nonumber\\
\nu_2&=&\frac{1}{3}\left(u+x+y-\sqrt{3}(x-y)\right),\nonumber
\eea 
\end{subequations} 
determine the sign of the partial transpose of 
$\tilde{\rho}_{AB}^{(j_1)}$. Related numerical results will be 
presented below.  

\subsubsection{Isotropic quantum channels}  
For $d> 3$ the number of independent parameters in the problem 
increases enormously with $d$, e.g. for $d=4$ we have 
\bea
\lambda_{mn}=\left (
\begin{array}{cccc}
\xi_0\quad & \eta_1\quad & \zeta\quad & \eta_1\\
\eta_1\quad & \eta_2\quad & \eta_3\quad & \eta_2 \\
\zeta\quad  & \eta_3\quad & \xi_1\quad & \eta_3\\
\eta_1\quad & \eta_2\quad & \eta_3\quad & \eta_2
\end{array}
\right ).
\label{lambdas4}
\eea
However, the situation becomes tractable in the case of isotropic 
channels (e.g., open-space QKD) where disturbances involving different 
errors \cite{noteD} are equal, thus leading to an eigenvalue matrix 
of the form \cite{PT,BKBGC,AGS}
\bea
\lambda_{mn}=\left (
\begin{array}{cccc}
u\quad & x\quad & \ldots\quad & x\\
x\quad & y\quad & \ldots\quad & y \\
\vdots\quad & \vdots\quad & \ddots\quad & \vdots\\
x\quad & y\quad & \ldots\quad & y 
\end{array}
\right ),
\label{lambdas-iso}
\eea 
for any dimension $d$.

In the case of qubits, such an isotropy argument does not seem to 
be a restriction. 
Thus, any eavesdropping strategy is equivalent to a 
symmetric (i.e., isotropic) one \cite{CG-FGNP}. 
This might be due to the fact that such a symmetry arises 
automatically as an inherent property of the qubit-based 
QKD protocols [the matrix (\ref{lambdas2}) is of the form 
(\ref{lambdas-iso})]. 
As a consequence, one is always able to substitute any eavesdropping 
attack with a symmetric one which yields the same results for all the 
properties which are defined as averages over all the possible 
messages sent by Alice to Bob (e.g., estimated disturbance) 
\cite{CG-FGNP}.   
Besides, here we see that the symmetry (isotropy) arises 
automatically for 
qutrits [the matrix (\ref{lambdas3}) is for the form 
(\ref{lambdas-iso})] and thus similar arguments must hold for 
$d=3$ as well.  
Nevertheless, we have found that for $d > 3$ this symmetry does not 
exist [see for instance Eq. (\ref{lambdas4}) for $d=4$] and one 
has to apply it explicitly. Hence, unless the quantum channel itself 
is isotropic, a restriction to symmetric eavesdropping strategies for 
$d>3$ seems unreasonable and might, in general, underestimate 
Eve's power. However, in our case such a restriction does not seem 
to affect the threshold disturbance, while simultaneously enabling 
us to present numerical results regarding $2d$-state QKD protocols 
with $d>3$. 

So, using the matrix (\ref{lambdas-iso}), the normalization 
condition (\ref{normG}) reads  
\bea
u+2(d-1)x+(d-1)^2y=1,\nonumber
\eea
and only two of the three parameters $(u,x,y)$ 
are independent. 
Moreover, combining Eqs. (\ref{Proj2}), (\ref{QBER2-2}) and 
(\ref{rhoAB-Bell-G}) we have that the average 
estimated disturbance is given by
\bea 
D = (d-1)x + (d-1)^2y.\nonumber
\eea 
Finally, in analogy to the case of qutrits, the partial transpose of 
${\tilde \rho}_{AB}^{(j_1)}$ is block-diagonal with each block being 
a $d\times d$ matrix. For odd dimensions all the blocks are identical 
whereas for even dimensions two different blocks appear.  

\subsubsection{Numerical results and discussion}
\begin{figure}
\centerline{\includegraphics[width=8.0cm]{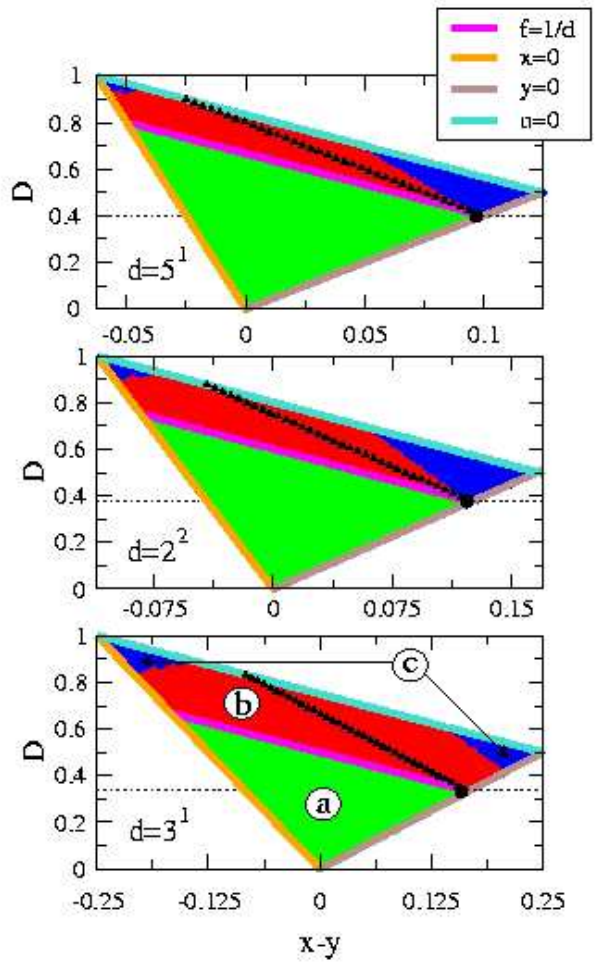}}
\caption{(Color online) $2d$-state QKD protocols: 
The regions of the independent parameters $D$ and $x-y$ for 
which the qudit-pair state $\tilde{\rho}_{AB}^{(j_1)}$ is 
(a) NPPT and distillable ; ~
(b) PPT; ~
(c) NPPT but the reduction criterion is satisfied. 
From the top to the bottom, the ``distillability maps'' correspond 
to $d=5,4\,{\rm and}\,3$, respectively.
The non-negativity of $x,y\,{\rm and}\,u$ (straight lines)  
defines the region of parameters where the protocols operate 
while the distillability condition (\ref{fidel}) separates 
distillable from non-distillable states. 
The threshold disturbances for entanglement distillation are 
indicated by black dots. The triangles correspond to the 
separable state (\ref{sep}). Note the different scales of the 
horizontal axis.}
\label{D-x-y:fig}
\end{figure}

We have been able to test the results of Sec.\ref{distil-1} 
numerically for qutrits, while for higher dimensions 
we had to resort to the assumption of isotropic quantum channels, to 
reduce the number of independent parameters in our simulations.   
More precisely, fixing two independent parameters, say $D$ and $x-y$, 
we evaluated all the remaining parameters $u$, $x$, $y$ which are 
consistent with all the constraints. Subsequently, for the parameters 
at hand we checked whether the distillability condition 
(\ref{distilConst1}) is fulfilled and 
whether the two-qudit state ${\tilde \rho}_{AB}^{(j_1)}$ has a 
non-positive partial transpose (NPPT). 
The corresponding ``distillability maps'' for $d=3,4,5$ are presented 
in Figs.~\ref{D-x-y:fig}.

Our simulations confirm the validity of    
\bea
D_{\rm th} =  \frac{d-1}{2d}
\nonumber 
\eea
as the ultimate robustness bound for $2d$-state  QKD protocols. 
More precisely, for $D < D_{\rm th}$ Alice and Bob share 
always freely entangled qudit pairs [regime (a) in 
Figs.~\ref{D-x-y:fig}]. 
On the contrary, for $D \geq D_{\rm th}$ we can identify two different 
regimes of parameters. The dominant regime (b) involves parameters 
which yield a ${\tilde \rho_{AB}}^{(j_1)}$ with positive 
partial transpose (PPT). These states 
can not be distilled and are either separable or bound entangled 
\cite{HHH,Letal}. 
Besides, we have the regime of parameters (c), for which  
${\tilde \rho_{AB}}^{(j_1)}$ has a NPPT but the reduction criterion 
is not violated. These states probably belong to the 
hypothetical set of bound entangled states with NPPT 
\cite{HH,Letal}. 
At this point, it could be argued that $D\geq D_{\rm th}$ is the 
regime of 
parameters where the ideas of Horodecki {\em et al.} might be applicable for 
the distillation of a secret key from bound entangled states 
\cite{HHHO}.  
To this end, however, Alice and Bob have to confirm whether the state 
they share is indeed bound entangled. Such an identification is only  
possible with the help of appropriate additional entanglement 
witnesses constructed from the polarization data of the raw 
key \cite{CLL-AG}. 

\section{Conclusions}
\label{conclusions} 
We have discussed the robustness of qudit-based QKD 
protocols that use two mutually unbiased bases, 
under the assumption of general coherent (joint) attacks. 
For $d=3$ (i.e., for qutrits), we have presented evidence of the fact 
that any eavesdropping strategy is equivalent to a symmetric one, 
while for higher dimensions this equivalence is no longer valid. 
 
The lowest possible disentanglement bound that an eavesdropper 
can saturate in the context of these cryptographic protocols scales 
with dimension as $(d-1)/2d$. 
Whenever Alice and Bob detect disturbances above $(d-1)/2d$, they 
are not able to infer whether their correlations originate from an 
entangled state or not, and the protocol must be aborted. 
On the contrary, if the detected disturbance is below $(d-1)/2d$, 
the two honest parties can be confident that they share free 
entanglement with high probability and the extraction of a secret 
key is, in principle, possible.

In particular, for the entanglement-based version of the
protocols such a secure key can be obtained after
applying an appropriate EPP which purifies the qudit pairs 
shared between Alice and Bob towards $\ket{\Psi_{00}}$ 
\cite{DEJ,BDSW,ADGJ,HH}.
Moreover, in view of the fundamental role of entanglement in 
secret key distribution \cite{CLL-AG}, the development of qudit-based 
prepare-and-measure schemes that can tolerate bit error rates up to 
$(d-1)/2d$ is also possible. 
For this purpose, however, the construction of new appropriate two-way 
EPPs, which are consistent with 
the associated prepare-and-measure schemes seems to be of vital 
importance. 
Our results generalize the results of \cite{AGS} to arbitrary 
coherent attacks and simultaneously answer (to some extent) many 
of the open issues raised in the concluding remarks of that paper.

Finally, it should be stressed that the disturbance thresholds 
we have obtained depend on the post-processing of the QKD 
protocol. In particular, they rely on the complete omission of 
those qudits of the raw key for which Alice and Bob measured in 
different bases. Furthermore, they also rely on the fact that Alice 
and Bob manipulate each qudit pair separately.
Under these conditions, we have demonstrated that the extraction 
of a secret-key from bound entangled states is impossible in the 
framework of qudit-based QKD protocols that use two mutually unbiased 
bases.

\section{Acknowledgments}
Stimulating discussions with
Markus Grassl and Antonio Acin are gratefully acknowledged.
This work is supported by 
the EU within the IP SECOQC.


\begin{thebibliography}{99}
\bibitem{note_prime}
However, it is worth noting that many of the subesequent arguments 
are also valid for dimensions which are not prime powers.
\bibitem{half}
It is worth noting that a logarithmic scaling of the size of the 
random sample with the length of Alice's and Bob's key, 
is sufficient for security issues. See Ref. \cite{chau} 
for a rigorous proof.
\bibitem{notation}
Clearly, the kets and bras here refer to the $j_i$-th check pair. 
In order not to overload notation, however, we have dropped the 
indices $j_i$ on the left-hand side of this equation.
\bibitem{noteD}
In general, as is clearly reflected in the projector (\ref{Proj1}),  
the qudit state $\ket{l_B}$ sent from Alice to Bob may 
undergo $d-1$ different ``dit-flip'' errors, i.e. 
$\ket{l_B}\to\ket{(l+k)_B}$ with $k\in\field{d}^*$.  
In particular, if $d$ is prime, then $k\in\{1,2,\ldots,d-1\}$.
\end{thebibliography}
\end{document}